\documentclass[aps,prl,twocolumn,groupedaddress]{revtex4-1}

\usepackage{amsmath}
\usepackage{amsfonts}
\usepackage{amssymb}
\usepackage{xspace}
\usepackage{graphicx}
\usepackage[utf8]{inputenc}
\usepackage[unicode]{hyperref}
\usepackage{color}
\usepackage{braket}
\usepackage{bbold}
\usepackage{diagbox}

\setcitestyle{square,numbers}

\newcommand{\ie}[0]{i.e.\@\xspace}
\newcommand{\eg}[0]{e.g.\@\xspace}

\renewcommand{\S}[0]{\mathcal{S}}

\newcommand{\fan}[1]{\hat{c}^{\vphantom\dagger}_{#1}}
\newcommand{\fcr}[1]{\hat{c}^{\dagger}_{#1}}

\newcommand{\ban}[1]{\hat{a}^{\vphantom\dagger}_{#1}}
\newcommand{\bcr}[1]{\hat{a}^{\dagger}_{#1}}

\newcommand{\rhoan}[1]{\hat{\rho}^{\vphantom\dagger}_{#1}}

\newcommand{\Bond}[1]{\hat{B}_{#1}}

\newcommand{\fcohan}[1]{c_{#1}}
\newcommand{\fcohcr}[1]{\bar{c}_{#1}}
\newcommand{\fcohmeasure}{\mathcal{D}(\bar{c},c)}

\newcommand{\rhocohan}[1]{\rho_{#1}}

\newcommand{\Pp}{P}

\newcommand{\omz}{\omega_0}

\newcommand{\kF}{k_{\text{F}}}

\newcommand{\Hc}{\mathrm{H.c.}}

\newcommand{\expv}[1]{\left\langle #1 \right\rangle}

\newcommand{\expvz}[1]{\left\langle #1 \right\rangle_0}

\begin{document}

\title{Directed-Loop Quantum Monte Carlo Method for Retarded Interactions}

\author{Manuel Weber}
\author{Fakher F. Assaad}
\author{Martin Hohenadler}

\affiliation{\mbox{Institut f\"ur Theoretische Physik und Astrophysik,
Universit\"at W\"urzburg, 97074 W\"urzburg, Germany}}

\date{\today}

\begin{abstract}
The directed-loop quantum Monte Carlo method is generalized to the case of
retarded interactions. Using the path integral, fermion-boson or spin-boson
models are mapped to actions with retarded interactions by analytically
integrating out the bosons. This yields an exact algorithm that combines the
highly efficient loop updates available in the stochastic series expansion
representation with the advantages of avoiding a direct sampling of the bosons.
The application to electron-phonon models reveals that the method overcomes
the previously detrimental issues of long autocorrelation times and
exponentially decreasing acceptance rates. For example, the resulting
dramatic speedup allows us to investigate the Peierls quantum phase
transition on chains of up to $1282$ sites.
\end{abstract}

\maketitle

{\it Introduction.}---In the absence of general exact solutions for strongly
correlated quantum systems, the development of efficient numerical methods  is
a central objective. For the one-dimensional (1D) case, the
density-matrix renormalization group (DMRG) method \cite{White92,White93}
has become the standard. However, it is much less efficient for higher dimensions,
finite temperatures, or long-range interactions, so that quantum Monte Carlo (QMC)
methods are often advantageous.  The latter yield high-precision results for
rather general 1D fermion and spin Hamiltonians. In particular, the cost for QMC simulations
of path integrals in the stochastic series expansion (SSE) representation
\cite{PhysRevB.43.5950} scales linearly with system size $L$ and inverse
temperature $\beta=1/k_\text{B} T$. Autocorrelation times are short due
to the use of cluster updates (operator loops or directed loops) \cite{PhysRevB.59.R14157,Sandvik02,PhysRevE.71.036706}.
While usually restricted to 1D fermionic models by the sign problem, such QMC methods were
successfully applied in higher dimensions to models of spins \cite{PhysRevLett.98.227202,Carrasquilla15,PhysRevLett.114.105303}
and bosons \cite{PhysRevLett.94.207202,PhysRevA.84.041608}.

Retarded interactions (\ie, nonlocal in imaginary time) impose significant limitations
regarding system size, temperature, and parameters. They typically arise from a coupling to bosonic
modes (\eg, phonons \cite{HOLSTEIN1959325,PhysRevLett.42.1698} or spin
fluctuations \cite{Edwards2006133}), or from dynamical screening
\cite{PhysRevB.1.471}. Because such problems are generically nonintegrable, much of
their understanding comes from numerical investigations. While
impurity problems can be solved very efficiently
\cite{PhysRevLett.97.076405}, lattice problems remain a challenge. 
In the DMRG \cite{PhysRevB.60.7950,0295-5075-84-5-57001,TeArAo05}, the
large bosonic Hilbert space becomes a limiting factor, especially for
finite-temperature or dynamical properties. For SSE-based QMC methods
\cite{ClHa05,hardikar:245103,PhysRevB.67.245103,PhysRevB.92.245132,PhysRevLett.83.195,PhysRevB.56.14510,arXiv:0705.0799},
the absence of global updates for the bosons  (except for a specific form of
spin-phonon coupling \cite{Assaad2008}) makes simulations
significantly less efficient than for fermions or spins. Autocorrelation
times increase strongly near phase transitions \cite{hardikar:245103}.
Extremely long autocorrelation times also affect determinant QMC
methods, where even the sampling of free bosons can be challenging
\cite{Hohenadler2008}. In fact, for
electron-phonon models, the continuous-time interaction-expansion (CT-INT) QMC
method \cite{PhysRevB.72.035122,PhysRevB.76.035116,Assaad14_rev} with a $(\beta L)^3$
scaling and local updates produces better results for correlators at the same
numerical cost \cite{PhysRevB.92.245132}. On the other hand, directed-loop methods
remain efficient for interactions that are long-ranged in space
\cite{PhysRevE.68.056701,PhysRevB.85.195115}.

In this Letter, we overcome these limitations by formulating the problem in
imaginary time, integrating out the bosons analytically, and using
directed-loop updates to efficiently sample the resulting problem with a
retarded interaction. This novel approach combines the advantages of global
updates available in the SSE representation and the analytical integration
over the bosons possible in the action-based CT-INT method.

{\it Method.}---The SSE representation \cite{PhysRevB.43.5950}
corresponds to a  high-temperature expansion of the partition function,
\begin{align}
\label{Eq:SSE_Ham}
 Z
  =
  \sum_{\alpha} \sum_{n=0}^{\infty} \frac{\beta^n}{n!} \sum_{S_n}
  \bra{\alpha} \prod_{p=1}^{n} \hat{H}_{a_p,b_p} \ket{\alpha} \, .
\end{align}
The Hamiltonian $\hat{H}$ is written as a sum of local
operators, $\hat{H}=- \sum_{a,b} \hat{H}_{a,b}$, where $a$ specifies an
operator type and $b$ the bond between sites $i(b)$ and $j(b)$. The
expansion~(\ref{Eq:SSE_Ham}) is sampled stochastically. A 
configuration with expansion order $n$ corresponds to a string of
$n$ operators, specified by the index sequence $S_n=\{ [a_1,b_1], \dots, [a_n,b_n] \}$, and a state $\ket{\alpha}$ from
a complete basis in which $\hat{H}$ is nonbranching, \ie, $\hat{H}_{a,b} \ket{\alpha}\sim\ket{\alpha'}$.
If $\hat{H}$ contains both off-diagonal ($a=1$) and
diagonal ($a=2$) operators, two types of updates are sufficient
to achieve ergodicity. For the {\it diagonal updates}, it is convenient to
fix the length of the operator string to $N$ by inserting $N-n$ unit
operators ($a=0$). Then, $S_N$ can be traversed sequentially and updates
$\hat{H}_{0,b}\leftrightarrow \hat{H}_{2,b}$ be proposed. The corresponding
configuration weights are directly obtained from the propagated state
$\ket{\alpha(p)}\sim \prod_{l=1}^p \hat{H}_{a_l,b_l}\ket{\alpha}$.
The global {\it directed-loop updates}
interchange diagonal and off-diagonal operators on an extensive number of
bonds \cite{Sandvik02}.

For concreteness, we explain our method for the highly nontrivial 1D spinless
Holstein Hamiltonian \cite{HOLSTEIN1959325}
\begin{equation}\label{eq:Holstein}
\hat{H}= -t \sum_{i} \Bond{i,i+1} + \omz \sum_{i} \bcr{i} \ban{i}
+ \gamma \sum_i \rhoan{i}  ( \bcr{i} + {\ban{i}})
\end{equation}
with the electronic hopping $\Bond{i,i+1} =
(\fcr{i} \fan{i+1} +\Hc)$ and the density $\rhoan{i} = (\fcr{i} \fan{i} -1/2)$; $\fcr{i}$ and $\bcr{i}$ are the usual fermionic and bosonic creation
operators acting at lattice site $i$; the chemical potential is zero. In the existing SSE approach to
fermion-boson models \cite{ClHa05,hardikar:245103,PhysRevB.67.245103,PhysRevB.92.245132}, the
fermions are sampled as explained above. Because of the absence of off-diagonal
terms, the bosons are updated by local moves with a cutoff 
for the local occupation number. Even using tempering, autocorrelation times increase
significantly with the coupling $\gamma$ \cite{hardikar:245103}, and acceptance rates decrease
exponentially for large $\omega_0$ \cite{PhysRevB.92.245132}, which severely
restricts applications.  Our new approach eliminates these problems.
It is based on the coherent-state
path integral \cite{Negele}, where
the Gaussian integrations over the bosonic fields are carried out
\cite{PhysRev.97.660} to obtain a fermionic action with
retarded interaction [we define $\lambda=\gamma^2/(2\omz t)$]
\begin{align}
\label{Eq:S_ret}
\S_{\text{ret}}
  =
  -2\lambda t \iint d\tau_1 d\tau_2 \sum_i \rhocohan{i}(\tau_1) \Pp(\tau_1-\tau_2) \rhocohan{i}(\tau_2)\,.
\end{align}
$\Pp(\tau) = \omz \cosh[\omz (\beta/2 - \tau)] / [2 \sinh(\omz\beta/2)]$
is the free boson propagator with $\tau\in[0,\beta)$ and $\Pp(\tau + \beta) = \Pp(\tau)$. Similar
interactions arise for other fermion-boson models.

In an action-based formulation, the SSE representation corresponds to an
expansion of $Z = \int
\fcohmeasure \,e^{-\S_0-\S_1}$ around $\S_0 = \int d\tau \sum_{i}
\fcohcr{i}(\tau) \, \partial_{\tau} \, \fcohan{i}(\tau)$. For a general action, we write $\S_1$ 
as a sum over vertices,
\begin{equation}\label{eq:S1sum}
  \S_1 = - \sum_{\nu} w_{\nu} h_{\nu}\,.
\end{equation}
A vertex is specified by a superindex $\nu$, a weight $w_{\nu}$, and the Grassmann
representation $h_{\nu}$ of an operator. The partition function becomes
\begin{align}
\label{Eq:Z_intexp}
Z
	=
		\sum_{n=0}^{\infty} \sum_{C_n} \frac{Z_0}{n!} \,
		w_{\nu_1} \dots w_{\nu_n}
		\expvz{ h_{\nu_1} \dots h_{\nu_n}}\,,
\end{align}
where $C_n = \{\nu_1, \dots, \nu_n\}$ encodes a configuration of order
$n$, $\expvz{O} = Z_0^{-1} \int \fcohmeasure \, e^{-\S_0} O$ with $Z_0 = \int
\fcohmeasure \, e^{-\S_0}$, and time-ordering is implicit. The 
expansion~(\ref{Eq:Z_intexp}) converges for any finite $\beta$ and $L$ \cite{PhysRevB.72.035122}.

For problems without retardation, we have $\S_1 = - \int d\tau \sum_{a,b} H_{a,b}(\tau)$, \ie, $\nu=\{a,b,\tau\}$,
$w_{\nu} = d\tau$, and $h_{\nu} = H_{a,b}(\tau)$. 
The relation between Eq.~(\ref{Eq:Z_intexp}) and the SSE representation is
established by mapping the time-ordered expectation value to an
operator string:
\begin{equation}
\sum_{S_n} Z_0  \expvz{ h_{\nu_1} \dots h_{\nu_n}}
= \sum_{S_n}\sum_{\alpha} \bra{\alpha} \prod_p \hat{H}_{a_p,b_p}  \ket{\alpha}\,.
\end{equation}
On the right-hand side, the time labels are obsolete. Therefore, the $\tau$-integrations
contained in $\sum_{C_n}$ can be carried out and give $\beta^n$, leading to 
Eq.~(\ref{Eq:SSE_Ham}). A mapping between the SSE and a time-ordered
interaction expansion in the full Hamiltonian was introduced in
Ref.~\cite{PhysRevB.56.14510}.

For {\it retarded interactions} such as Eq.~(\ref{Eq:S_ret}), $\S_1$ contains
in particular the interaction $\S_{\text{ret}}$; $\expvz{h_{\nu_1} \dots h_{\nu_n}}$ can still be mapped to an operator string
to calculate the weight of a configuration. However, the fact that the
weight $w_\nu$ depends on imaginary time demands an explicit sampling of the
$\tau$-integrals, as well as time-ordering of the fields. Since $\S_1$ 
consists of bilinears $\overline{c}(\tau)c(\tau)$, this reordering
does not change the sign of the configuration.

{\it Algorithm.}---We illustrate our algorithm for the spinless
Holstein model; other models (see below) only require minimal modifications. As the standard
directed-loop method is well documented~\cite{Sandvik02}, we focus on the
differences. For technical details see also the SM.

\begin{figure}[t]
\centering
\includegraphics[width=0.65\linewidth]{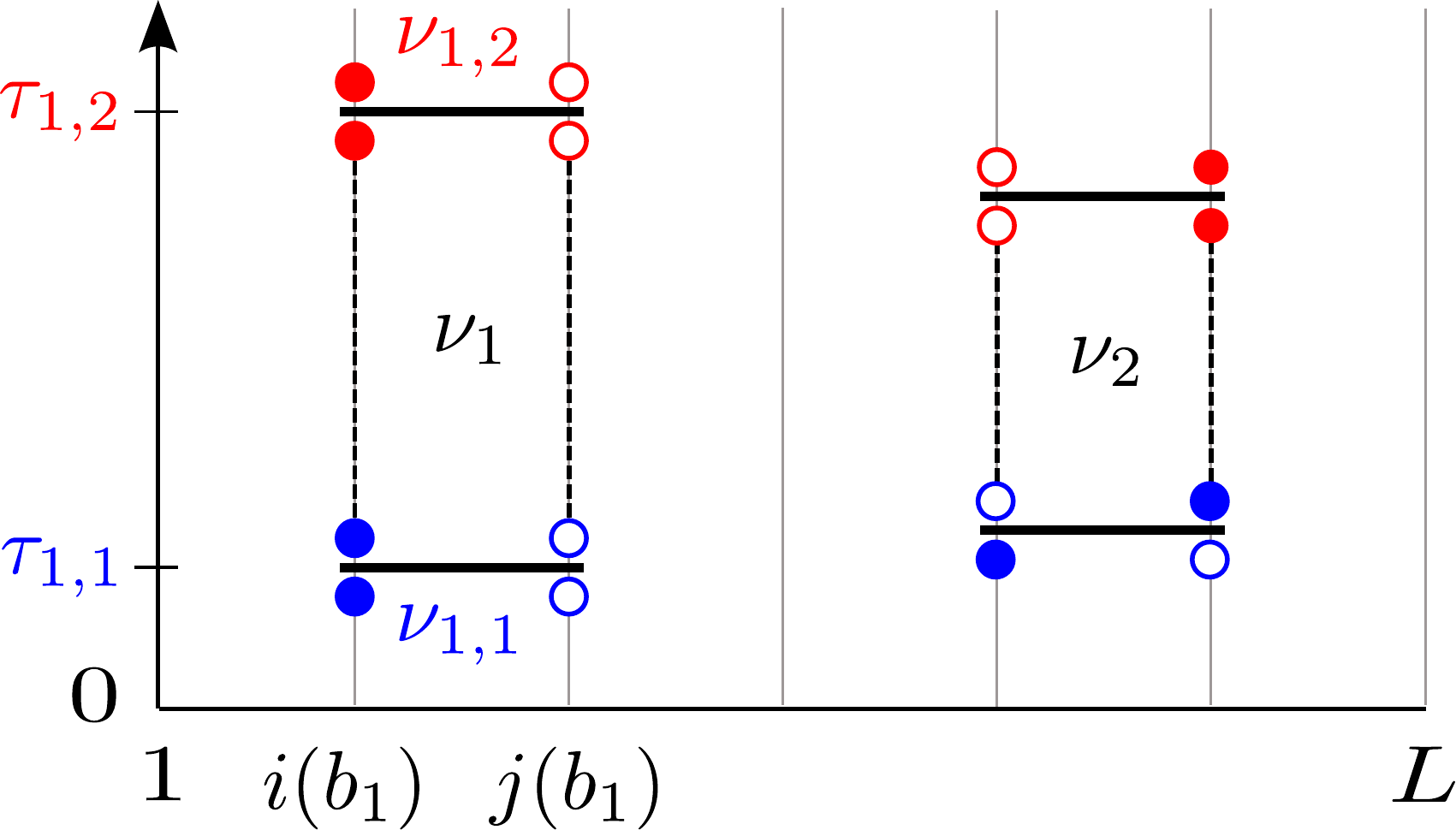}%
\caption{\label{fig:vertices} Vertices for the spinless Holstein model
[cf. Eqs.~(\ref{eq:vertices}) and~(\ref{eq:vertices2})]. Vertex $\nu_1$ is a diagonal vertex 
($a_{1,1}=a_{1,2}=2$) on bond $b_1$ connecting sites $i(b_1)$ and $j(b_1)$. It consists of subvertices $\nu_{1,1}$
at time $\tau_{1,1}$ and  $\nu_{1,2}$ at $\tau_{1,2}$. Vertex $\nu_2$ is
off-diagonal ($a_{2,1}=1$, $a_{2,2}=0$) and acts at $b_2$, $\tau_{2,1}$,
$\tau_{2,2}$. Open (solid) symbols indicate empty (occupied) lattice sites.
}
\end{figure}

{\it (i) Configuration space}.---A configuration consists of a state
$\ket{\alpha} = \ket{n_1,\dots,n_L}$ in the local occupation
number basis, an expansion order $n$, and an ordered vertex list
$C_n = \{ \nu_1, \dots, \nu_n \}$. For a coherent representation of
both the bilinear hopping terms with one time argument
and the biquadratic interaction terms with two time arguments,
it is expedient to write each vertex $\nu_k$ as two
subvertices $\nu_{k,1}$ and $\nu_{k,2}$ and add a unit ``operator''
$\mathbb{1}_{b}(\tau)$ with a dummy time variable to the hopping terms. Each subvertex then has local variables $\nu_{k,j} = \{ a_{k,j},
b_{k,j}, \tau_{k,j} \}$ (see Fig.~\ref{fig:vertices}). For the Holstein model, $b_{k,1} = b_{k,2} = b_k$.
To lighten the notation we drop the index $k$ from here on.
We write Eq.~(\ref{eq:S1sum}) as
\begin{align}\label{eq:S1vert}
\S_1 = - \iint d\tau_1 d\tau_2 \, \Pp(\tau_1-\tau_2) \sum_{a_1, a_2, b} h_{a_1a_2,b}(\tau_1,\tau_2) \, .
\end{align}
The off-diagonal hopping vertices are given by 
\begin{align}\label{eq:vertices} 
h_{10,b}(\tau_1,\tau_2)
	&=
	\frac{t}{2} B_{b}(\tau_1) \mathbb{1}_{b}(\tau_2) \, ,
\\\nonumber
h_{01,b}(\tau_1,\tau_2)
	&=
	\frac{t}{2} \mathbb{1}_{b}(\tau_1)  B_{b}(\tau_2) \,,
\end{align}
whereas the diagonal interaction vertices read
\begin{align}\label{eq:vertices2} 
h_{22,b}(\tau_1,\tau_2)
  &=
    \lambda t \left[ C + \rhocohan{i(b)}(\tau_1) \rhocohan{i(b)}(\tau_2) 
    + (i\leftrightarrow j) \right] 
\end{align}
with $j(b)=i(b)+1$. To arrive at the form~(\ref{eq:S1vert}),
we multiplied the off-diagonal terms in Eq.~(\ref{eq:vertices}) with the
bosonic propagator and exploited $\int_0^{\beta} d\tau_2 \, \Pp(\tau_1-\tau_2)=1$
for the dummy time variables. This essentially promotes the hopping terms to
retarded interactions and yields a vertex weight $\mathcal{W}_\nu=w(\tau_1,\tau_2) \, W[h_{a_1a_2,b}(\tau_1,\tau_2)]$.
Here, $w(\tau_1,\tau_2) = \Pp(\tau_1 -\tau_2) \, d\tau_1 d\tau_2$ irrespective of the operator types $a_1$, $a_2$.
 As a result, $P(\tau_1-\tau_2)$
only plays a role for the diagonal updates but drops out of the directed-loop
equations, allowing for a simple and efficient implementation.
In contrast,
$W[h_{a_1a_2,b}(\tau_1,\tau_2)]$ depends on time only implicitly via the
world line configuration
and its values are given in the SM.
Finally, the constant $C=1/2 + \delta$ ($\delta \ge 0$) in
Eq.~(\ref{eq:vertices2}) ensures positive weights.

{\it (ii) Diagonal updates}.---For retarded interactions, the operator string cannot be traversed sequentially because
each vertex update requires knowledge of the propagated state 
at two distinct positions in the string. 
However, the occupation number at $\{i,\tau\}$ is completely determined
by the initial state $\ket{\alpha}$ and the number of off-diagonal operators
that act between $0$ and $\tau$ and involve site $i$.
During the diagonal updates, we construct an ordered list containing
the time arguments of the operators $B_{b(i)}(\tau)$ for each $i$. Sorting this list takes $\mathcal{O}(L\beta \log \beta)$
operations, after which any propagated state can be quickly calculated. Diagonal updates involve adding or removing a single vertex $h_{22,b}(\tau_1,\tau_2)$
using the Metropolis-Hastings algorithm \cite{1953JChPh..21.1087M,10.2307/2334940} with acceptance rates
$
A_{C \to C'}
=
\min \left( R_{C\to C'}, 1 \right)
$.
For the addition of a new vertex  
$R_{C_n \to C_{n+1}}
=
{L\beta  W[h_{22,b}(\tau_1,\tau_2)]}/{(n_{\text{diag}}+1)}
$, whereas for the removal $R_{C_n \to C_{n-1}} = 1/R_{C_{n-1} \to C_{n}}$.
Here, $n_{\text{diag}}$ is the number of diagonal vertices in $C_n$.
Sampling $\tau_1$, $\tau_2$ according to $\Pp(\tau_1-\tau_2)$ by inverse
transform sampling ensures high acceptance rates for any $\omega_0$.

{\it (iii) Directed-loop updates}.--Directed-loop updates are very similar
for retarded and instantaneous interactions \cite{Sandvik02}. In the latter case, $\ket{\alpha(p)}$
is updated along a closed path connecting a subset of the vertices. Starting
at a leg $l_\text{i}$ of a randomly chosen vertex, the choice of the exit leg
$\l_\text{e}$ determines how the vertex changes as a result of the flipping of
the occupation numbers $n_{l_\text{i}},n_{l_\text{e}}$ to $1-n_{l_\text{i}},1-n_{l_\text{e}}$.
Thereby, the operator type of the vertex can change from $a=1$ to $a=2$
or vice versa. From $\l_\text{e}$ the loop continues to
the next vertex until it closes. The probabilities for choosing $l_\text{e}$ are determined
by the directed-loop equations for a general vertex \cite{Sandvik02}, which
can be derived from the requirement of local detailed balance. 

Our generalization to retarded interactions exploits (i) the
  subvertex structure introduced above, (ii) the fact that the update of a subvertex only
changes the world line configuration locally into another allowed
configuration, and (iii) our choice of the weight $w(\tau_1,\tau_2)$
that removes any time dependence from the directed-loop equations. Because of
(i) and (ii) each subvertex becomes an independent entry to the usual
linked vertex list \cite{Sandvik02} that also includes the unit
operators. While (ii) allows us to update subvertices individually, the retarded
interaction~(\ref{Eq:S_ret}) leads to an update probability that also 
depends on the other subvertex connected via $P(\tau)$. These conditions hold for
the Holstein model (see SM) but also for, \eg, Fröhlich,
Su-Schrieffer-Heeger, and spin-phonon models \cite{MHHF2017}.
Finally, for the spinless Holstein model, the directed-loop equations can be solved
exactly and backtracking is absent for $\lambda\leq 1$ (see SM).

{\it (iv) Observables}.---Electronic observables are calculated exactly as in
the SSE representation \cite{Sandvik91}. Bosonic estimators are obtained
using generating functionals
\cite{PhysRevB.94.245138}. Dynamic correlation functions
are also accessible.

\begin{figure}[t]
\centering
\includegraphics[width=0.9\linewidth]{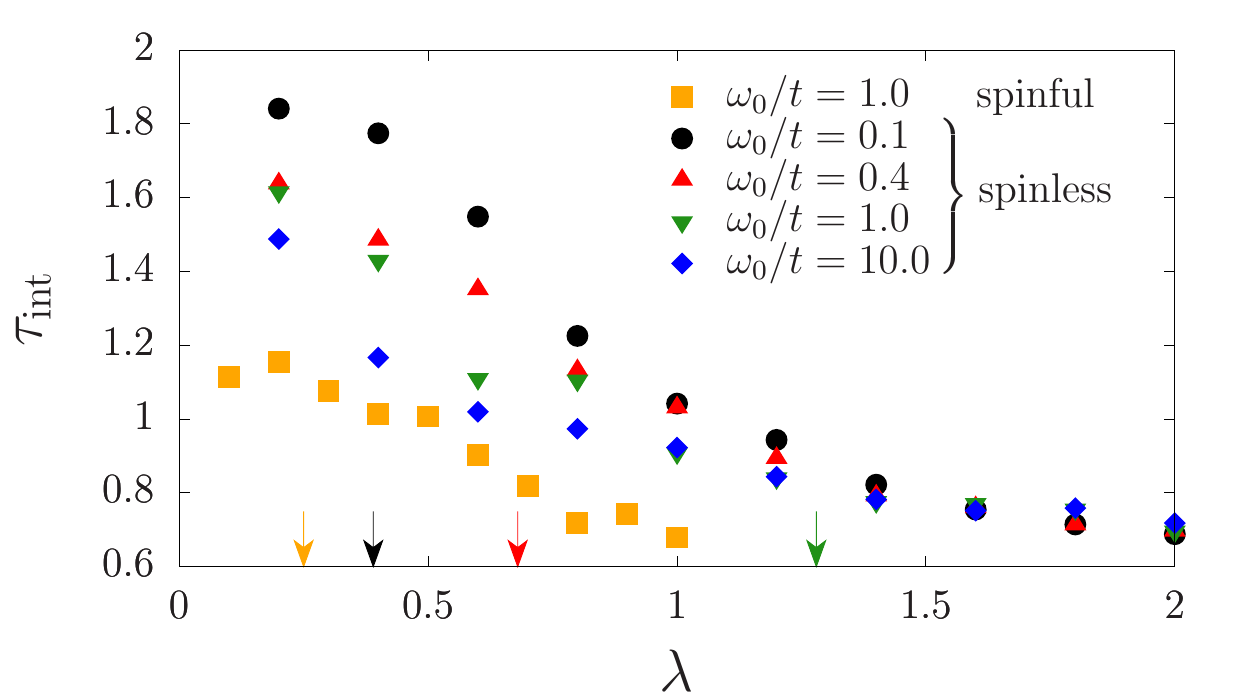}%
\caption{\label{fig:tauint}
Autocorrelation time $\tau_\text{int}$ for the total energy, as
determined from a rebinning analysis \cite{Janke2008}, for the spinless and
the spinful Holstein model. Here, $L=18$, $\beta t=2L$. Arrows indicate Peierls
critical values $\lambda_c(\omega_0)$~\cite{0295-5075-84-5-57001,0295-5075-87-2-27001}.
}
\end{figure}

{\it Application}.---To demonstrate the potential of our new method, we first
discuss its efficiency. In standard SSE simulations of the 
Holstein-Hubbard model---the spinful analog of Eq.~(\ref{eq:Holstein})---the
integrated autocorrelation time $\tau_\text{int}$ essentially diverges with
$\lambda$ \cite{hardikar:245103}. Although reduced by parallel tempering,
$\tau_\text{int}$ exceeds $100$ at intermediate coupling already for
moderately difficult parameters ($\omega_0=t$, $L=16$, $\beta t=2L$) \cite{hardikar:245103}.
Figure~\ref{fig:tauint} shows $\tau_\text{int}$ for our method for $L=18$ and $\beta t=2L$
\footnote{{A sweep consisted of two blocks of diagonal and directed-loop
  updates. For each block of diagonal updates, we attempted approximately
$2\expv{n_{\text{diag}}}$ updates. The number of loop updates was fixed by
touching approximately $2\expv{n}$ subvertices of type $a=1,2$.}}, covering
the entire range of phonon frequencies
from adiabatic to antiadiabatic and the entire range of couplings from weak
to strong. Periodic boundary conditions were used.
Remarkably, $\tau_\text{int}$ is of order 1 both for the spinful and the spinless Holstein model
\footnote{For the spinful Holstein model, each subvertex obtains an additional
spin variable $\sigma_j$. Including this variable for the dummy unit ``operators'' in
the off-diagonal vertices leads to an additional prefactor of $1/2$.}.
Autocorrelations in fact decrease with increasing $\lambda$, with no
visible signature of the Peierls phase transition. The data shown are for the
total energy, 
for other observables 
$\tau_\text{int}$ is even smaller. Similar
autocorrelation times were observed for larger systems.

Having established its numerical efficiency, we used the method to
obtain high-precision results for the half-filled spinless Holstein model~(\ref{eq:Holstein}).
The latter provides a generic framework to study the Peierls
transition of 1D electrons coupled to quantum phonons. From previous work
\cite{PhysRevB.27.4302,PhysRevLett.80.5607,PhysRevB.58.13526,PhysRevB.73.245120,0295-5075-87-2-27001},
the model is known to exhibit a Berezinskii–Kosterlitz–Thouless quantum phase
transition with dynamical exponent $z=1$ between a Luttinger liquid and a
charge-density-wave (CDW) insulator with a $q=2\kF=\pi$ modulation of charge
density and lattice deformations. Since $z=1$ we keep $\beta/L=\text{const}$.

\begin{figure}[t]
\centering
\includegraphics[width=\linewidth]{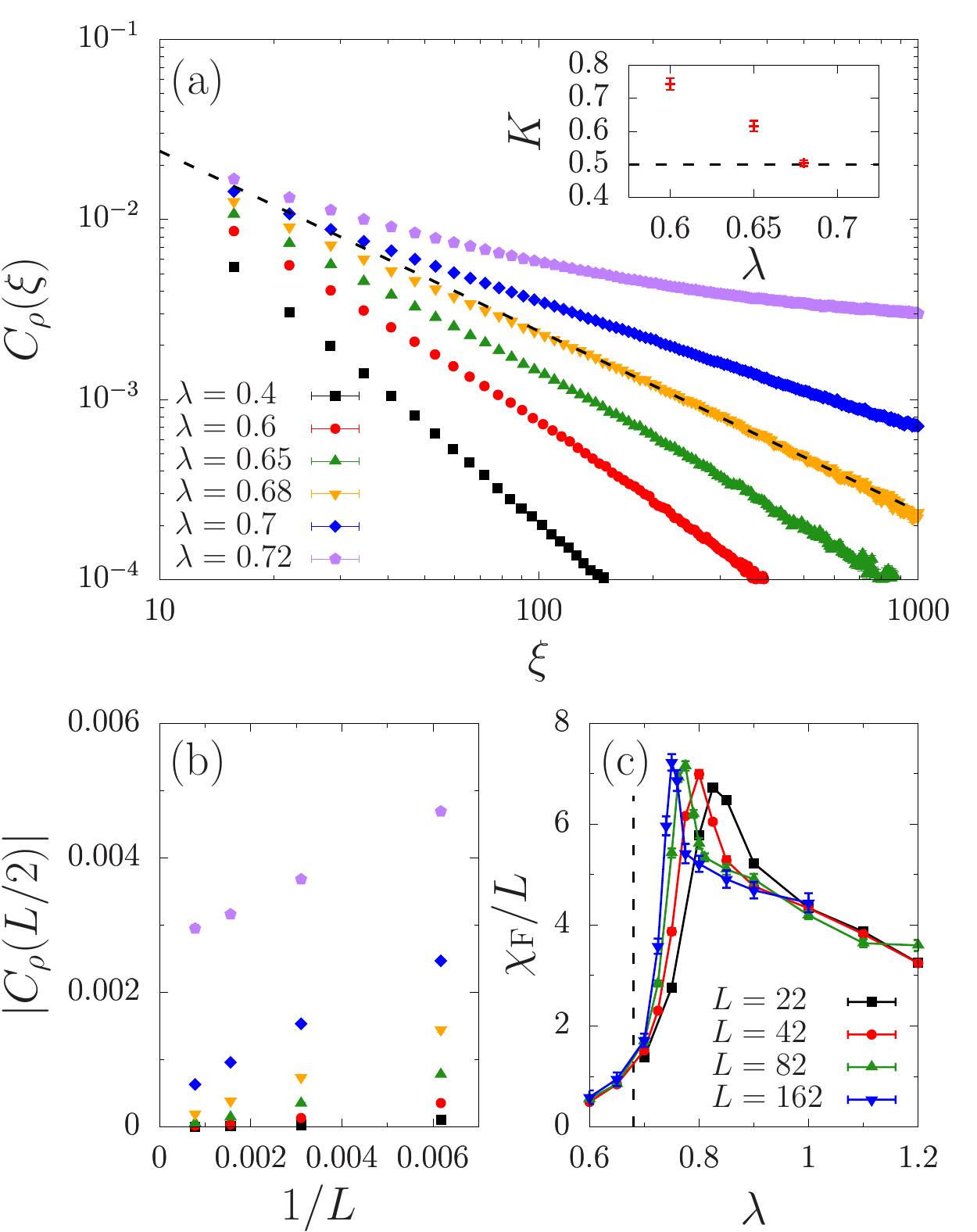}%
\caption{\label{fig:PT} Results for the spinless Holstein model ($\omega_0=0.4t$).
(a) Real-space density correlator for even distances as a function of the conformal distance
$\xi = L\sin (\frac{\pi r}{L})$ \cite{Cardy1996} on chains of up to $L=1282$ sites ($\beta t=2L$).
The dashed line indicates the $1/\xi$ decay expected at $\lambda_c$.
Inset: Luttinger parameter $K$ extracted from fits of $C_\rho(L/2)$ to
$a/r^{2K}$ using $L=162$\,--\,562.
(b) Finite-size scaling of the density correlations at distance $L/2$,
indicating long-range order beyond $\lambda_c=0.68(1)$. Here, $\beta t=2L$ and
the key is the same as in (a).
(c) Fidelity susceptibility for $\beta t=4L$. The dashed line indicates
$\lambda_c$.
}
\end{figure}

Figure~\ref{fig:PT} shows the real-space density correlator $C_{\rho}(r) = \langle
\hat{\rho}_r \hat{\rho}_0 \rangle$ (using the conformal distance $\xi$, see caption)
and the fidelity susceptibility $\chi_\text{F}$ \cite{PhysRevX.5.031007,PhysRevB.94.245138},
a finite-temperature extension of the quantum fidelity and an unbiased diagnostic for quantum phase transitions \cite{PhysRevE.74.031123,2008arXiv0811.3127G}. We simulated
systems of up to $L=1282$ sites with $\beta t\geq 2L$. Real-space correlation
functions were previously reported for $L\lesssim 50$ \cite{PhysRevB.92.245132}, DMRG results for other
quantities were available up to $L=256$ \cite{0295-5075-87-2-27001}.

Figure~\ref{fig:PT}(a) reveals the theoretically predicted power-law decay of
density correlations in a spinless, repulsive Luttinger
liquid \cite{Voit94}. The dominant contribution to $C_\rho(r)$ is the
oscillating term $\cos(2\kF r) \, r^{-2K}$ (we only plot even $r$). 
The nonuniversal exponent is determined by the Luttinger
parameter $K$. As expected for the Mott transition of a spinless Luttinger
liquid, $K$ decreases with increasing $\lambda$ until it reaches the critical
value $K=1/2$ for $\lambda_c=0.68(1)$ \cite{0295-5075-87-2-27001}. This can be seen by comparing to the
dashed line in Fig.~\ref{fig:PT}(a) that shows a $1/r$ power law. The inset
shows estimates for $K$ from power-law fits (see caption for details).
For $\lambda>\lambda_c$, $K$ scales to zero and the system exhibits long-range
CDW order. 

In Fig.~\ref{fig:PT}(b), we plot the density correlator at the largest
distance $r=L/2$, whose thermodynamic limit serves as an order parameter for
the quantum phase transition. We find a nonzero extrapolated order parameter for
$\lambda\gtrsim0.68$, in accordance with Fig.~\ref{fig:PT}(a) and
previous estimates \cite{0295-5075-87-2-27001}.
The transition can also be detected from the
fidelity susceptibility shown in Fig.~\ref{fig:PT}(c). Because statistical
errors are generally larger for $\chi_\text{F}$, the maximum system size was $L=162$.
In contrast to previous work, the directed-loop algorithm permits us to reach
sufficiently large values of $L$ and $\beta$ to observe the cusp at
$\lambda_c$ predicted theoretically \cite{PhysRevB.91.014418}. The latter
sharpens and converges (slowly, similar to other 1D models \cite{PhysRevB.91.014418}) to 
$\lambda_c$ with increasing $L$. More generally, Figs.~\ref{fig:PT}(b) and~\ref{fig:PT}(c) are important because they establish the usefulness of the order parameter
and $\chi_\text{F}$ to detect the CDW transition without reference to
bosonization results. They can therefore be used for spinful electron-phonon
models, the analysis of which is complicated by the existence of a spin gap
in the metallic phase \cite{Lu.Em.74,Voit98,PhysRevB.92.245132} and the
absence of a reliable theory for the Mott transition of a Luther-Emery liquid \cite{Lu.Em.74}.
Moreover, our method can access the system sizes necessary to resolve
the spin gap \cite{PhysRevB.92.245132}.

{\it Conclusions and Outlook}.---We have introduced a highly efficient
directed-loop QMC method for systems with retarded interactions. 
For the electron-phonon models considered, our algorithm outperforms any
other existing method, including the DMRG. Because of the global updates,
autocorrelations are negligible and there is no need for tempering or machine learning.
The method permits us to study fermion-boson models with the same
accuracy as purely fermionic models. It can be used to solve a number
of open problems in the field of electron-phonon
physics. These include the phase diagrams of models with competing
electron-phonon and electron-electron interactions, the specific
heat of quantum Peierls chains, and the dimensional crossover as a
function of temperature in Peierls materials.
The method can also be extended to spin-boson models via a fermionic
path-integral representation with a suitable constraint. Sign-free
simulations of spins or hardcore bosons with retarded interactions
can be carried out in any dimension and are important for the
understanding of correlated quantum systems with dissipation (for recent
numerical work see, \eg, Refs.~\cite{PhysRevLett.113.260403,1704.00606}).
Finally, it will be interesting to explore if the method permits us to study quasi-1D
materials beyond the low-energy regime by including higher bands via a
frequency-dependent interaction.

{\begin{acknowledgments}%
    This work was supported by the German Research Foundation (DFG) through SFB
    1170 ToCoTronics and FOR~1807. The authors gratefully acknowledge the computing time granted by the John
    von Neumann Institute for Computing (NIC) and provided on the supercomputer
    JURECA \cite{Juelich} at the J\"{u}lich Supercomputing Centre.
\end{acknowledgments}}


%

\pagebreak

\section{Supplemental Material}

In the following, we provide details on our implementation of the
directed-loop algorithm for the half-filled spinless Holstein model that may be helpful
for readers wishing to implement the method themselves or modify an existing
directed-loop code.

\subsection{Vertex weights}

\begin{figure}[b]
\centering
\includegraphics[width=\linewidth]{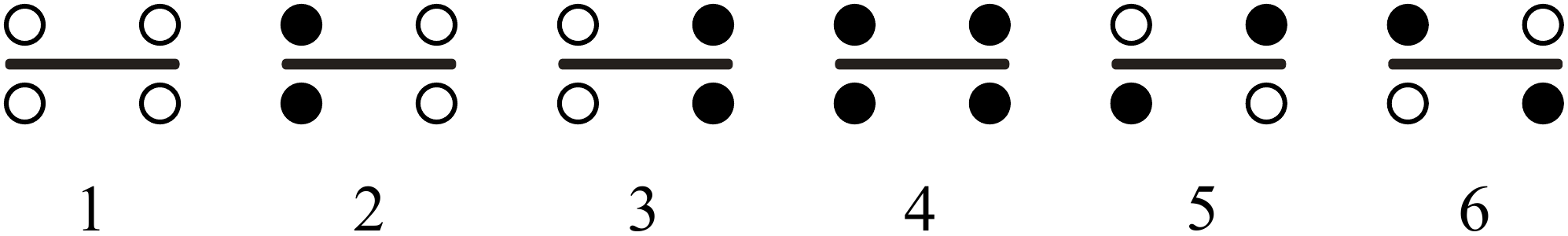}%
\caption{\label{fig:vertextypes}
Subvertex types for the spinless Holstein model.
Open (filled) symbols indicate empty (occupied) lattice sites.
}
\end{figure}

In the SSE representation, the weight of a Monte Carlo configuration,
$
W(C_n)
	=
		\frac{1}{n!}
		\prod_{p=1}^n 
		\mathcal{W}_{\nu_p}
$,
factorizes into a product of individual vertex weights $\mathcal{W}_\nu=w(\tau_1,\tau_2) \,  W[h_{a_1a_2,b}(\tau_1,\tau_2)]$.
The explicit time dependence of the vertex is in
$w(\tau_1,\tau_2) = \Pp(\tau_1 -\tau_2) \, d\tau_1 d\tau_2$, which is
independent of the operator type and therefore has to be considered only during
the diagonal updates. The remainder $W[h_{a_1a_2,b}(\tau_1,\tau_2)] = W_{v_1,v_2}$ is fully determined by
the vertex types $v_1,v_2\in \{1,\dots,6\}$ that in turn specify the 
change of the world-line configuration at each subvertex.
Figure~\ref{fig:vertextypes} shows the possible subvertex types
for the spinless Holstein model, where $v\in  \{1,\dots,4\}$ corresponds to
unit and diagonal operators ($a=0,2$) and $v\in  \{5,6\}$ to
off-diagonal ones ($a=1$). The corresponding weights $W_{v_1,v_2}$ are
given in Table~\ref{Tab:weights}.

 \begin{table}[htbp]
 \caption{\label{Tab:weights}
Vertex weights $W_{v_1,v_2}$ for the spinless Holstein model for all possible
combinations of vertex types $v_1$ and $v_2$.
}
 \begin{ruledtabular}
 \renewcommand{\arraystretch}{1.2}
 \begin{tabular}{c|cccccc}
\diagbox{$v_1$}{$v_2$}   & 1 & 2 &3 &4 &5 &6 \\
\hline
1 & $\lambda t \left(C +\frac{1}{2}\right)$ & $\lambda t C$ & $\lambda t C $ & $\lambda t \left(C -\frac{1}{2}\right)$ & $t/2$ & $t/2$ \\
2 & $\lambda t C$                      & $\lambda t \left(C +\frac{1}{2}\right)$ & $\lambda t \left(C -\frac{1}{2}\right) $ & $\lambda t C $ & $t/2$ & $t/2$ \\
3 & $\lambda t C $                     & $\lambda t \left(C -\frac{1}{2}\right)$ & $\lambda t \left(C +\frac{1}{2}\right) $ & $\lambda t C$ & $t/2$ & $t/2$ \\
4 & $\lambda t \left(C -\frac{1}{2}\right)$  & $\lambda t C$ & $\lambda t C $ & $\lambda t \left(C +\frac{1}{2}\right)$ & $t/2$ & $t/2$ \\
5 & $t/2$                                     & $t/2$ & $t/2$ & $t/2$ & $0$ & $0$ \\
6 & $t/2$                                     & $t/2$ & $t/2$ & $t/2$ & $0$ & $0$ \\
 \end{tabular}
 \end{ruledtabular}
 \end{table}

\subsection{Solution of the directed-loop equations}

\begin{figure}[t]
\centering
\includegraphics[width=\linewidth]{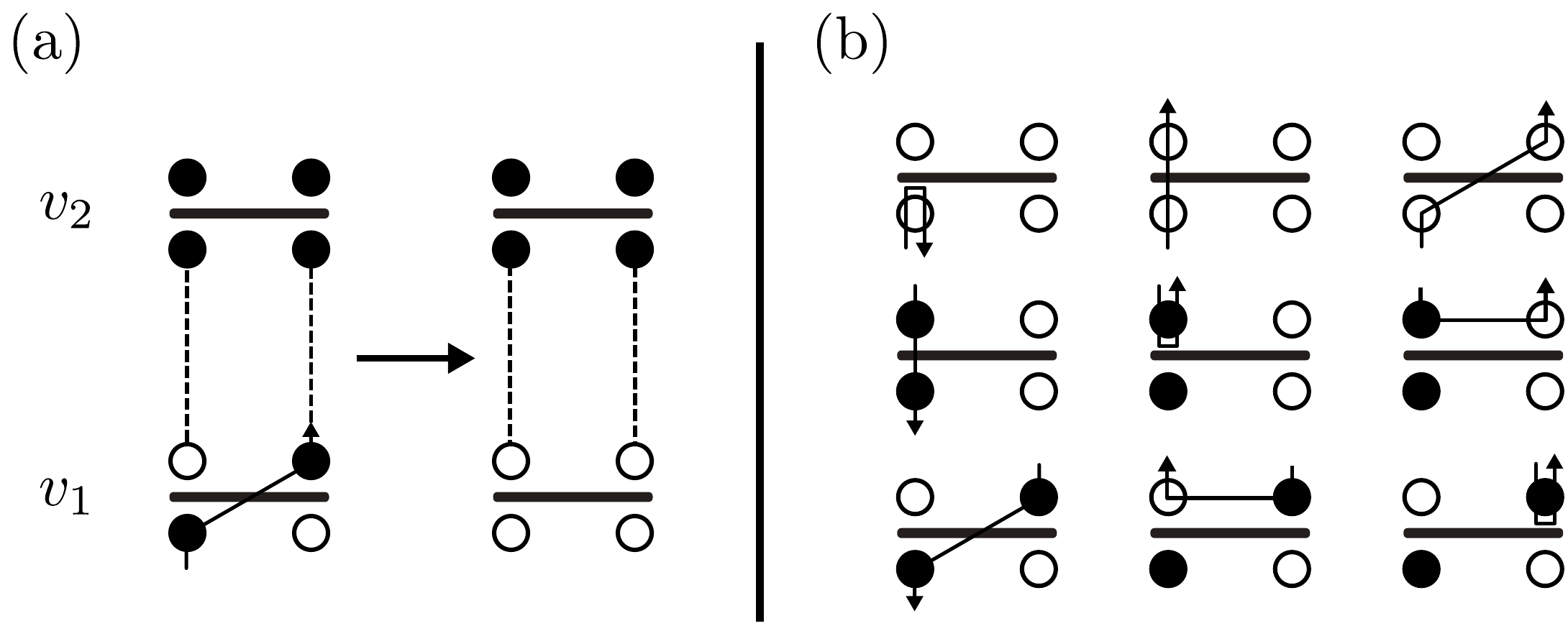}%
\caption{\label{fig:assignments}
(a) A vertex is determined by the subvertex types $v_1$ and $v_2$.
The directed path is only assigned to one subvertex and flips the occupation numbers
of the corresponding states. Here, we consider $h_{10,b}(\tau_1,\tau_2) \to h_{22,b}(\tau_1,\tau_2)$. (b) Example of an assignment table for the
directed-loop equations for a subvertex of type $v_1$.
}
\end{figure}

For the directed-loop updates, the configuration space of vertex types
$v\in \{1,\dots,6\}$ shown in Fig.~\ref{fig:vertextypes} is enlarged by
assigning to a given vertex directed paths that connect an entrance leg
$l_\text{i} \in  \{1,\dots,4\}$ with an exit leg $l_\text{e} \in  \{1,\dots,4\}$. The Monte
Carlo weights of these assignments are determined by the directed-loop
equations, which can be derived from the requirement of local detailed balance
\cite{Sandvik02}. For retarded interactions, each vertex consists of two subvertices.
While the directed-loop equations
determine the weight of the total vertex, the loop is constructed locally by only
assigning a directed path to one subvertex and leaving the other unchanged.
As shown in Fig.~\ref{fig:assignments}(a), the occupation numbers on the sites
included in the loop are then switched and the vertex type changes.
For the vertices defined by Eqs.~(8) and (9) of the main text, we have to distinguish two
cases: the directed loop either hits a unit operator, or any other operator.
For the former case, the path goes straight through the subvertex with
probability $1$ and changes its vertex type.
For the latter case, the directed-loop equations have to be solved explicitly.

As discussed in the main text, the world-line configuration is updated at
each subvertex independently. However, while the vertex type of the other
subvertex does not change, the same is not in general true for its
dummy operator type $a_2$. For example, $a_2$ changes from $0$ to $2$ when 
interchanging $h_{10,b}(\tau_1,\tau_2) \leftrightarrow h_{22,b}(\tau_1,\tau_2)$
in Fig.~\ref{fig:assignments}(a). This corresponds to the update from a
hopping operator at $\tau_1$ and a unit operator at $\tau_2$ to a (diagonal)
density-density interaction term at times $\tau_1$ and $\tau_2$. Because unit
operators that change into diagonal operators are relevant for the weights
$W_{v_1,v_2}$ in later updates, it is important to keep track of such changes.

We illustrate the solution of the directed-loop equations for the assignment table
given in Fig.~\ref{fig:assignments}(b) (for a detailed discussion of assignment
tables see Ref.~\cite{Sandvik02}). We only show the possible assignments
for vertex type $v_1$, corresponding to the lower subvertex in Fig.~\ref{fig:assignments}(a).
The second subvertex of type $v_2$ remains unaffected by this segment of the
loop update but still enters the configuration weight.
Each row in Fig.~\ref{fig:assignments}(b) shows the possible assignments for 
a fixed $l_{\text{i}}$ and the three possible exit legs $l_{\text{e}}$.  The
associated weights are symmetric around the diagonal because the corresponding 
assignments are related by  inverting the direction of the path and flipping
the occupation numbers on the sites touched by the loop. 
For the specific example of Fig.~\ref{fig:assignments}(b), we obtain for the corresponding weights
\begin{align}\nonumber
b_1 + a + b = W_{1,v_2} \, , \\
a + b_2 + c = W_{2,v_2} \, , \\\nonumber
b + c + b_3 = W_{5,v_2} \, .
\end{align}
The bounce weights $b_i$, $i\in \{1,2,3\}$, are related to the assignments
on the diagonal, whereas $a$, $b$, and $c$ are the remaining weights.
Our goal is to reduce the bounce weights and solve for $a$, $b$, and $c$.
To this end,  we write
\begin{align}\nonumber
a &= \frac{1}{2} \left[ W_{1,v_2} + W_{2,v_2} - W_{5,v_2} - b_1 - b_2 + b_3 \right] \, , \\
b &= \frac{1}{2} \left[ W_{1,v_2} - W_{2,v_2} + W_{5,v_2} - b_1 + b_2 - b_3 \right] \, , \\\nonumber
c &= \frac{1}{2} \left[ -W_{1,v_2} + W_{2,v_2} + W_{5,v_2} + b_1 - b_2 - b_3 \right] \, .
\end{align}
For concreteness, we choose $v_2=3$ and insert the weights given in
Table~\ref{Tab:weights}. This leads to
\begin{align}\nonumber
a &= \frac{1}{2} \left[ 2\lambda t C - \frac{(1+\lambda)t}{2}  - b_1 - b_2 + b_3 \right] \, , \\
b &= \frac{1}{2} \left[ \frac{(1+\lambda)t}{2} - b_1 + b_2 - b_3 \right] \, , \\\nonumber
c &= \frac{1}{2} \left[ \frac{(1-\lambda)t}{2} + b_1 - b_2 - b_3 \right] \, .
\end{align}
The bounce weights $b_i$ and the constant $C=1/2 +\delta$ must be chosen such 
that $a$, $b$, and $c$ are positive. For $\lambda < 1$, this is already fulfilled by
$b_1=b_2=b_3=0$ and $\delta \geq (1-\lambda)/(4\lambda)$. In our simulations,
we have chosen the lower bound.
For $\lambda \geq 1$, the positivity of $c$ requires $b_1 \geq (\lambda-1)t/2$, whereas
the positivity of $b$ demands $b_1 \leq (\lambda+1)t/2$. We have chosen the lower bound and $\delta=0$.
This procedure has to be repeated for each type of background vertex $v_2$ and each possible
assignment table for $v_1$. In the end, we find that the global constant $C$ has
to be chosen as for the example given here.

Let us point out that exact solutions of the directed loop equations are neither common
nor necessary for efficient simulations. Instead, the equations can be solved using
linear programming techniques \cite{PhysRevE.71.036706}.

\end{document}